\documentclass[preprint,12pt]{elsarticle}
\usepackage[utf8]{inputenc}
\usepackage{graphicx}
\usepackage{bm}%
\usepackage{siunitx}
\usepackage{textcomp}
\usepackage{lineno}
\usepackage{xcolor}
\journal{NIMA}
\begin{document}
\begin{frontmatter}



\title{Statistical Image Reconstruction for High-Throughput Thermal Neutron Computed Tomography}


\author{J.~M.~C.~Brown$^{1,2}$, U.~Garbe$^3$ and D.~Pelliccia$^4$}
\address{$^1$ Department of Radiation Science and Technology, Delft University of Technology, The Netherlands \\
         $^2$ Centre for Medical Radiation Physics, University of Wollongong, Australia \\
         $^3$ Australian Centre for Neutron Scattering, Australian Nuclear Science and Technology Organisation, Australia \\
         $^4$ Instruments \& Data Tools Pty Ltd, Australia
}

\begin{abstract}
Neutron Computed Tomography (CT) is an increasingly utilised non-destructive analysis tool in material science, palaeontology, and cultural heritage. With the development of new neutron imaging facilities (such as DINGO, ANSTO, Australia) new opportunities arise to maximise their performance through the implementation of statistically driven image reconstruction methods which have yet to see wide scale application in neutron transmission tomography. This work outlines the implementation of a convex algorithm statistical image reconstruction framework applicable to the geometry of most neutron tomography instruments with the aim of obtaining similar imaging quality to conventional ramp filtered back-projection via the inverse Radon transform, but using a lower number of measured projections to increase object throughput. Through comparison of the output of these two frameworks using a tomographic scan of a known 3 material cylindrical phantom obtain with the DINGO neutron radiography instrument (ANSTO, Australia), this work illustrates the advantages of statistical image reconstruction techniques over conventional filter back-projection. It was found that the statistical image reconstruction framework was capable of obtaining image estimates of similar quality with respect to filtered back-projection using only 12.5\% the number of projections, potentially increasing object throughput at neutron imaging facilities such as DINGO eight-fold.

\end{abstract}


\end{frontmatter}
\section{Introduction}
Neutron Computed Tomography (CT) is an increasingly utilised non-destructive analysis tool in material science, palaeontology, and cultural heritage \cite{Lehmann2005,Schwarz2005,Kardjilov2011,DINGO1}. It is able to simultaneously provide unique elemental and structural information about the interior of these objects with information that may be complementary to that of X-ray based approaches \cite{Schwarz2005,Vontobel2005,Vontobel2006}. With the development of new neutron imaging facilities (such as DINGO, ANSTO, Australia \cite{DINGO1}) new opportunities arise to maximise their performance through the implementation of statistically driven image reconstruction methods which have yet to see wide scale application in neutron transmission tomography \cite{Anderson2009,Kardjilov2018}. This class of reconstruction methods are almost exclusively used for X-ray CT in the clinical setting \cite{Beister2012,Geyer2015}, and have begun to filter through to other X-ray imaging research tools (i.e. synchrotrons and coherent x-ray lab sources \cite{Hahn2015,Hehn2018}) which, until recently, were dominated by algebraic methods \cite{Mayo2003,Paganin2006,Gureyev2006,Beltran2010,Nesterets2015,Vagovic2015}. 

Statistical Image Reconstruction (SIR) approaches possess a number of advantages over algebraic approaches for statistically sparse imaging methodologies like neutron CT \cite{Lange1,Fessler2000}. The main motivation of this research is the desire to increase sample throughput whilst maintain imaging performance. Measurement of tomographic scans at neutron imaging instruments can take several hours, and in some instances several days, depending on the object and imaging task due to the low neutron flux \cite{Anderson2009,Pfeiffer2006}. In addition irradiation times on this order of magnitude can result in activation of specific metals within each sample above the radiation safety action thresholds requiring their isolation until they return to a safe level \cite{Anderson2009}. 

This work outlines the implementation of a Convex Algorithm Statistical Image Reconstruction (CA-SIR) framework \cite{Lange1} applicable to the geometry of most neutron imaging instruments with the aim of obtaining similar imaging quality to the most commonly used algebraic Image Reconstruction (IR) approach, Ramp Filtered Back-Projection via the inverse Radon transform (R-FBP) \cite{Kardjilov2011,Anderson2009,Kardjilov2018}, from a lower number of measured projections to increase object throughput. These two frameworks were applied to a tomographic scan of a known phantom obtained with the neutron radiography instrument DINGO at the OPAL research reactor (ANSTO, Australia) \cite{DINGO1} and their recovered object reconstructions compared. Section \ref{section:M} describes the experimental procedure used to obtain the tomographic data, developed CA-SIR framework, benchmarking with respect to R-FBP and discusses the image assessment metrics. The results, their discussion, and an overall conclusion then follow in Section \ref{section:M}, \ref{section:D} and \ref{section:C} respectively.

\section{Method}
\label{section:M}
\subsection{Acquisition of Tomographic Data}

\begin{figure}[tb]    
 \centering 
 \includegraphics[width=0.7\textwidth]{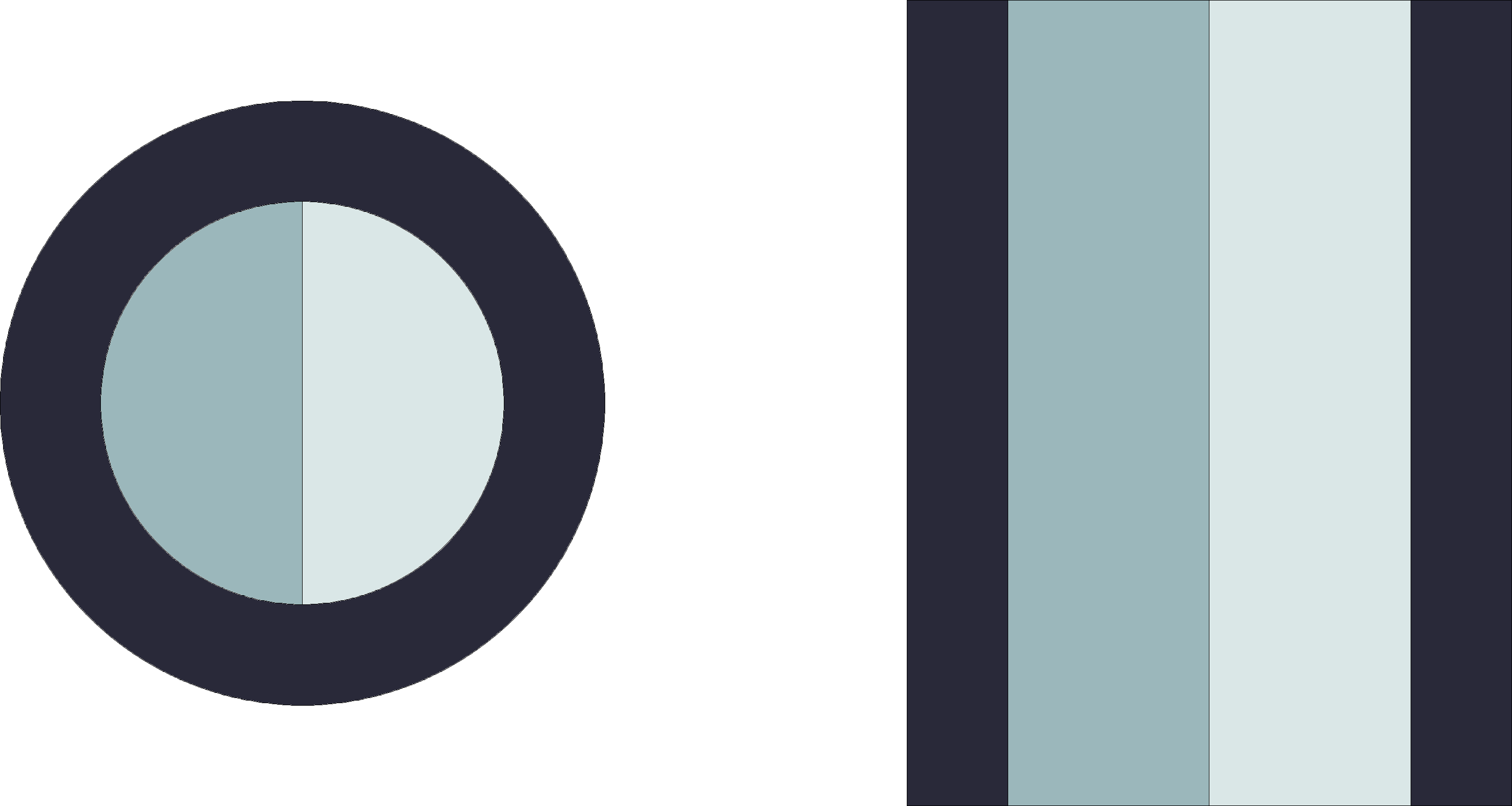}
\caption{The cylindrical phantom employed to assess the performance of the develop Convex Algorithm Statistical Image Reconstruction (CA-SIR) framework. Inner regions are composed of Titanium (Ti) (left) and Aluminium (Al) (right) half cylinders of 10 mm in radius and 40 mm height, with a Stainless Steel (S.S.) outer shell 5 mm thick and 40 mm in height. The colour scale used represents the three materials linear attenuation coefficient at the maximum intensity of DINGOs Maxwell-Boltzmann distributed thermal spectrum (1.5 \si{\angstrom}): S.S. 1.131 cm$^{-1}$, Al 0.101 cm$^{-1}$, and Ti 0.450 cm$^{-1}$ \cite{Sears,Munter}.}
\label{fig:1}
\end{figure}

A cylindrical phantom composed of two half rods of Aluminium (Al) and Titanium (Ti) inserted inside a Stainless Steel (S.S.) tube (see Figure \ref{fig:1}) was selected as the piece-wise uniform sample to assess the performance of the CA-SIR framework. The inner Al and Ti inserts are half cylinders of 10 mm in radius and 40 mm height, and the S.S outer shell is 5 mm thick with matching height. Its measurement was carried out with the neutron radiography instrument DINGO at the OPAL research reactor (ANSTO, Australia) in its high resolution configuration \cite{DINGO1}. In this configuration the neutron beam possesses a divergence on the order of 1 mrad, flux of 1.1 $\times 10^{7}$ n/(cm$^{2}$s), field-size of 50 $\times$ 50 mm$^{2}$ at the rotational sample stage, and Maxwell-Boltzmann distributed thermal spectrum with a maximum intensity at 1.5 \si{\angstrom} \cite{DINGO1}. The distance between the beam divergence defining pinhole and primary beam collimators was 2.5 m, and helium filled flight tubes were used to propagate the beam another 7.3 m to the rotational sample stage. A tomographic scan of 720 evenly spaced projections over 180 degrees was obtained with an exposure time of 30 s for each image. References images including a set of the direct beam for flat-field and background images (no beam) were also obtained resulting in a total measurement time over 9 hours.

The neutron detector located at the contact plane down stream of the rotational sample stage (e.g. within 1 cm to create a ``parallel beam" irradiation geometry) was composed of a scintillation screen, mirror and 2048 $\times$ 2048 pixel CCD camera (Andor IKON-L) mounted in a light tight housing \cite{DINGO1}. Within the light tight housing the scintillation screen, a 100 $\times$ 100 mm$^{2}$ Aluminium sheet coated with a 50 \textmu m thick layer of ZnS/6LiF, was centred and orientated perpendicular to the normal direction of the neutron beam. The mirror, located directly after the scintillation screen, is orientated at 45$^{\circ}$ to reflect the emitted light towards the CCD camera mounted at 90$^{\circ}$ above its centre, effectively removing it from the neutron beam path \cite{DINGO1}. The CCD camera was focused at the scintillation screen centre and images of 53.25 $\times$ 53.25 mm obtained giving an effective pixel size of 26 $\times$ 26 \textmu m.

\subsection{Convex Algorithm Statistical Image Reconstruction (CA-SIR) Framework}

Statistical image reconstruction methods differ from the deterministic reconstruction methods (i.e. FBP, ART) through the inclusion of counting statistics during the reconstruction process \cite{Beister2012,Fessler2000}. In the case of transmission tomography, the key assumption of these methods is that the statistical variation in signal that occurs from the particle flux propagation through an object and then measured at the detection plane can be described via a Poisson distribution \cite{Fessler2000}. With an appropriate set of parameters/process that describes the particle flux propagation through matter and measurement at the signal detection plane, it becomes possible to calculate the likelihood that this signal corresponds to an estimate of the object structure. Maximisation of the log of this likelihood will enable the estimation of the most probable object composition and structure \cite{Beister2012}. In this work the CA-SIR algorithm was selected over other statistical image reconstruction frameworks, such as classical Maximum Likelihood Expectation Maximisation (ML-EM), due to its accelerated convergence for transmission tomography \cite{Lange1}. Its general form, that updates the imaging volumes attenuation coefficient estimates $\mu^{n}$ with each step $n+1$, can be written as:

\begin{equation}
	 \displaystyle \mu^{n+1}_{j} = \mu^{n}_{j} + \frac{\mu^{n}_{j}\sum_{i}l_{ij}\left[d_{i} e^{-\left< l_{i},\mu^{n}\right>} - Y_{i}  \right]}{\sum_{i}l_{ij}\left< l_{i},\mu^{n}\right>d_{i} e^{-\left< l_{i},\mu^{n}\right>}} \label{eqn1}        
	\end{equation} 

\noindent with $\left< l_{i},\mu^{n}\right> = \sum_{j}l_{ij}\mu^{n}_{j}$ and where:

\begin{itemize}
 \item $\mu_{j}$ is the neutron attenuation coefficient at position $j$ (represented by voxel $j$),
 \item $Y_{i}$ is the measured value at $i$, where $i$ indicates both the vector of the line projection ($l_{i}$) and its index within the detector array,
 \item $l_{ij}$ is the contribution of voxel $j$ to the line projection $i$ normalised as a function of the total number of projections,
 \item $d_{i}$ is the flat-field intensity value of the detector at $i$, and
 \item $d_{i} e^{-\left< l_{i},\mu^{n}\right>}$ represents is the forward projected estimate at the pixel $i$ at the surface of the detector array.
\end{itemize}
\noindent Further information on the CA-SIR algorithm, and discussion of its validity for application in transmission tomography, can be found in Lange and Fessler \cite{Lange1}.

A direct implementation of Siddon's line projection algorithm \cite{Siddon} was employed as the basis forward and back projection steps. Both the pinhole-detector image blurring and detector noise were included within the forward projection model. Here the pinhole-detector image blurring was modelled as a single 1D Gaussian of 78 \textmu m Full Width at Half Maximum (FWHM) that was applied to each line of the forward projected sinogram. The detector noise was modelled as a Poisson distribution, approximated as a Gaussian distribution, and its FWHM was calculated on a slice by slice basis from the obtained sets of flat-field and background images. 

\subsection{Comparison of Imaging Performance}

A total of 100 central object slices of the obtained tomographic data-set were reconstructed with both the CA-SIR and R-FBP framework\footnote{The R-FBP framework is based on a modified version of the inverse Radon algorithm contained in the scikit-image python library \cite{VanDerWalt}.}. Before reconstruction each sinogram was processed with a ring removal algorithm to suppress ring artefacts as is typically done in R-FBP \cite{Lyckegaard}. A single set of R-FBP images were reconstructed using all 720 projects. Whereas with the CA-SIR framework, seven sets of images were reconstructed using a total of 15, 30, 45, 90, 180, 360 and 720 evenly space projections over the sampled 180 degree tomographic data-set using 1000 iterations. Each set of 100 images were then analysed using the Figures of Merit (FoMs) defined below, with their average and standard deviation reported for comparison.

Three FoMs were selected to assess the quality of the recovered images of the cylindrical phantom: Signal to Noise Ratio (SNR), Contrast and Line Spread Function (LSF). The definition of each of these FoMs and their application to the recovered images follows below.

Signal to Noise Ratio (SNR) quantifies the noise of a feature in an image. The SNR is defined as:

\begin{equation}
 \displaystyle \text{SNR} = \frac{\bar{x}}{s}  \label{eqn2} 
\end{equation}

\noindent where $\bar{x}$ and $s$ are the mean and standard deviation of the pixel values in the feature respectively. The SNR was determined for all three material regions of the phantom.

Contrast quantifies the ability to differentiate materials. The Contrast is defined as:

\begin{equation}
 \displaystyle \text{Contrast} = \frac{|\bar{x}_{1}-\bar{x}_{2}|}{|\bar{x}_{1}+\bar{x}_{2}|}  \label{eqn3} 
\end{equation}

\noindent again $\bar{x}$ denotes the mean pixel value of each material under comparison. The contrast between materials was determined for all combinations within the phantom and image background.

Finally, the Line Spread Function (LSF) quantifies the image resolution at different material interfaces. Here, the LSF is given by:

\begin{equation}
 \displaystyle \text{LSF} = \frac{d (\text{ESF})}{dr} \label{eqn4} 
\end{equation}

\noindent which is the derivative with respect to distance $r$ of each material interfaces Edge Spread Function ($\text{ESF}$). The FWHM of the LSF was determined for all material interfaces with 50 horizontal line-profiles per recovered image ($\pm$ 25 lines around the object centre), including the outer S.S. shell of the phantom with image background.

%
%

\section{Results}
\label{section:R}
\begin{figure}[tb]
  \centering
  \includegraphics[width=1\textwidth, trim = {45 0 45 0},clip]{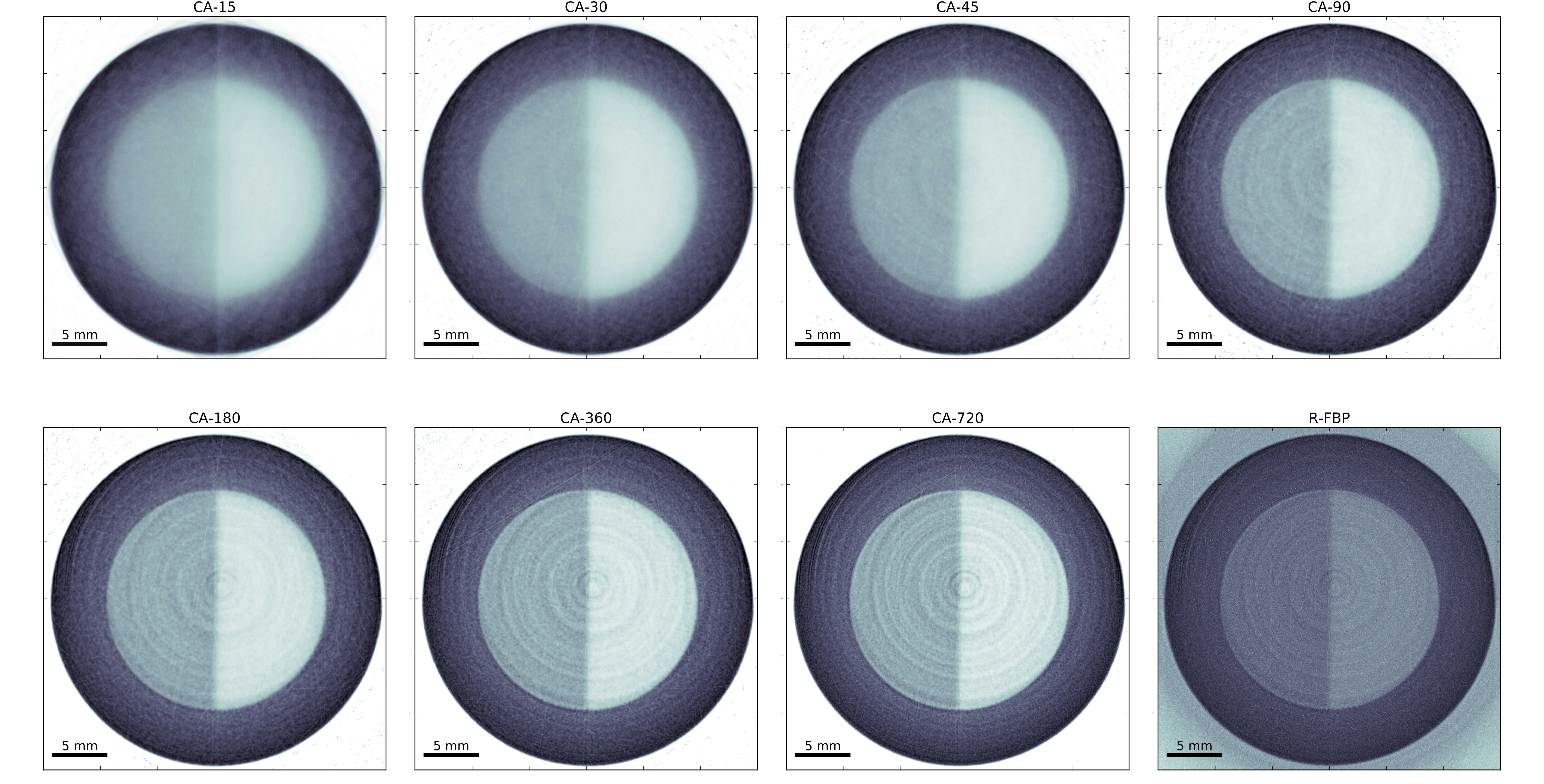}
  \caption{Recovered object neutron linear attenuation coefficient estimate slices from the centre of the imaged phantom for the R-FBP and CA-SIR different projection number trials of 15, 30, 45, 90, 180, 360 and 720 evenly space projections. Here each recovered slice's colour scale spans from zero to the maximum recovered linear attenuation coefficient.}
  \label{fig:2}
\end{figure}

A set of recovered object neutron linear attenuation coefficient estimate slices from the centre of the imaged phantom for the R-FBP and CA-SIR different projection number trials of 15, 30, 45, 90, 180, 360 and 720 evenly space projections sampled from the 180 degree tomographic data-set is presented in Figure \ref{fig:2}. Inspection of these recovered object estimates show that all three different regions of the phantom (S.S. outer sleeve, Ti half cylinder on the left and Al half cylinder on the right) can be distinguishes regardless of the number of projections utilised via the CA-SIR framework. Qualitative comparison of these images shows that: 1) the impact of ring artifacts in the CA-SIR images scales with projection number to rival that seen in the R-FBP image, and 2) the CA-SIR framework is able to recovered an object estimate of similar quality with respect to the R-FBP data with one-eighth the number of projections (i.e. CA-90 vs R-FBP). This observed correlation between impact of ring artifacts in the CA-SIR images and projection number can be contributed to an increase level of information from dead/hot pixels within the measured neutron detector data which mimics a material-interface like structure within the object \cite{Bushburg}.

\begin{figure}[tb]    
 \centering 
 \includegraphics[width=0.75\textwidth, trim = {25 0 55 35},clip]{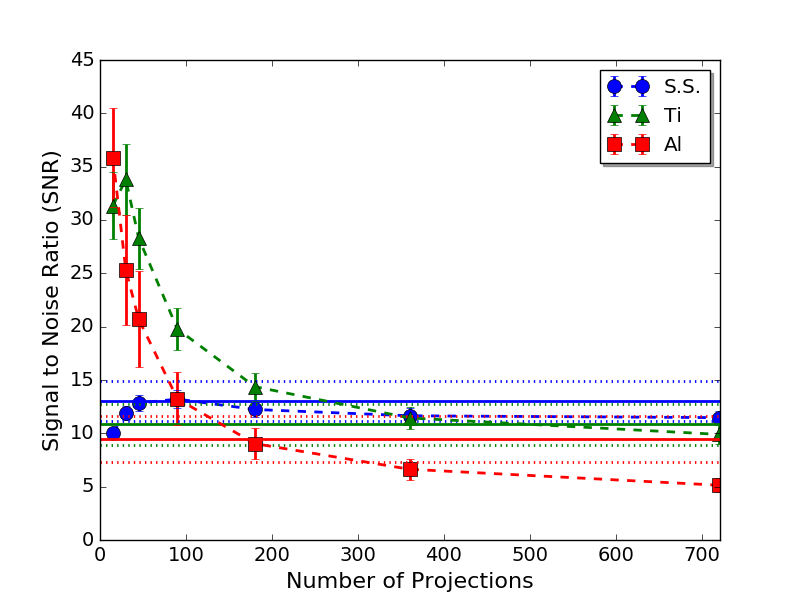}
\caption{Mean Signal to Noise Ratios (SNRs) of the 100 recovered image estimates for the S.S., Ti and Al regions of the imaged phantom. The R-FBP SNR mean and standard deviation values are represented as the solid and dotted horizontal lines respectively. Material region specific mean (markers) and standard deviation (error bars) SNR values obtained for each set of CA-SIR projection number trials has been overlaid.}
\label{fig:3}
\end{figure}

Figure \ref{fig:3} presents the SNR mean and standard deviation values of the three recovered phantom material regions within both the R-FBP and CA-SIR different projection number trials. Here the R-FBP SNR mean and standard deviation values are represented as the solid and dotted horizontal lines respectively for each material. Whereas the material region specific mean (markers) and standard deviation (error bars) SNR values obtained for each set of CA-SIR projection number trials has been overlaid in matching colour. Comparison of these data-sets shows that the SNR of S.S. and Ti regions obtained with CA-SIR is always within a standard deviation or better than the mean SNR of R-FBP. In the case of the Al region this is also true up until 360 projections with CA-SIR and then it falls below the low standard deviation band shortly afterwards. However a common trend of the CA-SIR trials for all three materials is that they exhibit an inverse correlation between projection number and SNR. 

\begin{figure}[tb]    
 \centering 
 \includegraphics[width=0.75\textwidth, trim = {25 0 55 35},clip]{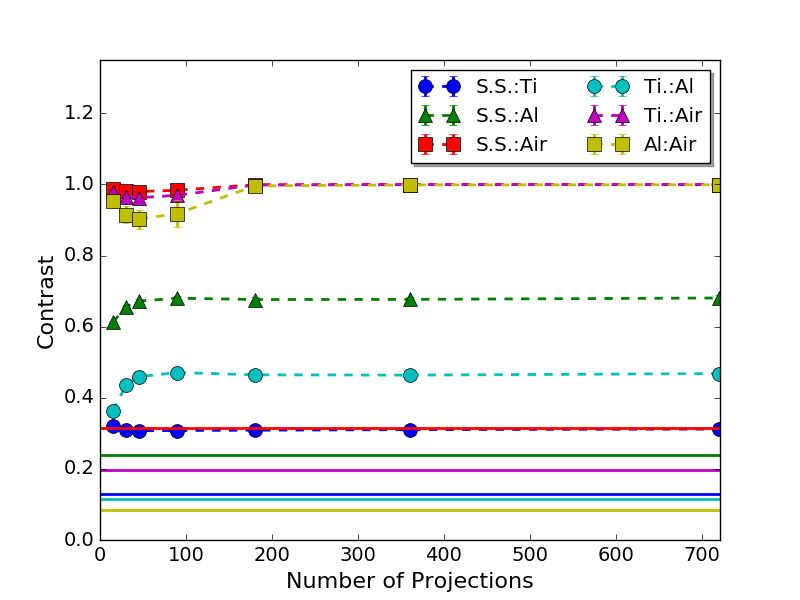}
\caption{Mean and standard deviation contrast values of the 100 recovered image estimates for the S.S., Ti, Al and air regions with respect to one another. Here the R-FBP mean contrast values are represented as the solid lines (the standard deviations are small and cannot be resolved with respect to their mean value lines). Whereas the mean (markers) and standard deviation (error bars) contrast values obtained for each set of CA-SIR projection number trials has been overlaid.}
\label{fig:4}
\end{figure}

Further quantification between R-FBP and the CA-SIR framework performance is presented in Figure \ref{fig:4} through comparison of the contrast between the recovered phantom material estimates and image background (air). It can be seen in this figure that: 1) in every case the contrast between features in all CA-SIR recovered image sets is higher than R-FBP, and 2) the contrast between features in the CA-SIR images plateaus at 90 projections. In particular the contrast between each of the phantom materials and the image background (air) in the CA-SIR case more closely matches expect values as their thermal neutron linear attenuation coefficient are at least two order of magnitude higher than air \cite{Sears,Munter}. This is also true for the cases of the contrast between S.S. and Ti with respect to Al as their linear attenuation coefficients differs via approximately an order of magnitude, and S.S. to Ti which differs via approximately a third for a monochromatic 1.5 \si{\angstrom} neutron beam \cite{Sears,Munter}.

\begin{figure}[tb]    
 \centering 
 \includegraphics[width=0.75\textwidth, trim = {25 0 55 35},clip]{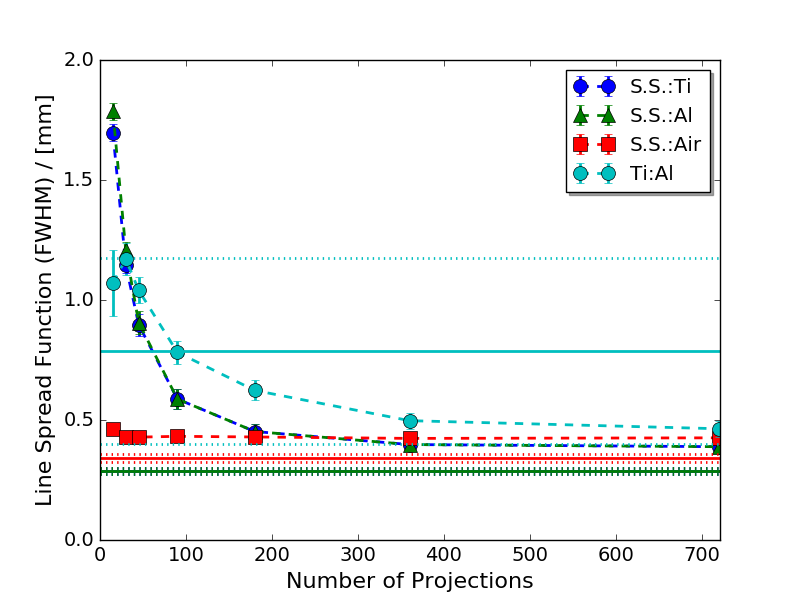}
\caption{Mean Line Spread Function (LSF) of all 5000 line profiles from the 100 recovered image estimates for interfaces between S.S., Ti, Al and air regions. The R-FBP LSF mean and standard deviation values are represented as the solid and dotted horizontal lines respectively. Material region specific mean (markers) and standard deviation (error bars) LSF values obtained for each set of CA-SIR projection number trials has been overlaid.}
\label{fig:5}
\end{figure}

Finally, the mean and standard deviation of the LSFs between each of the three phantom materials, and S.S. to Air, interfaces for both the R-FBP and CA-SIR different projection number trials can be seen in Figure \ref{fig:5}. Here the solid and dashed lines form the R-FBP data for the S.S. to Al interface (green) overlap the S.S. to Ti (blue) making them difficult to resolve. Also within this figure it can be seen that regardless of the number of projections used the CA-SIR frameworks' mean LSFs of all three S.S. material interfaces never come within the standard deviation of the R-FBP mean LSFs. Whereas in the case of the Ti to Al material interface the CA-SIR frameworks recovered LSFs are always within a standard deviation of the R-FBP mean LSFs, with a clearer definition between this material interface (indicated by a lower LSF) for projection number trials of 90 and above. Again all four material interfaces possess the common trend of CA-SIR trials exhibiting an inverse correlation between projection number and LSF.

\section{Discussion}
\label{section:D}
Both visual qualitative and quantitative assessment of the performance of the CA-SIR framework for different projection number trials of 15, 30, 45, 90, 180, 360 and 720 with respect to the commonly used R-FBP for neutron tomography was undertaken for a three material cylindrical phantom at the neutron radiography instrument DINGO (ANSTO, Australia). With the exception of the LSF for the S.S. to other material interfaces, the CA-SIR framework is shown able to recover comparable or better than object estimates then R-FBP for projection number trials down to 90 in Figures \ref{fig:1} through \ref{fig:5}. Based on these qualitative and quantitative findings it can be stated that the CA-SIR framework is capable of obtaining image estimates of similar quality with respect to R-FBP using only 12.5\% the number of object projections, potentially increasing object throughput at neutron imaging facilities such as DINGO, ANSTO, Australia eight-fold. 

The differences in LSF for the S.S. to other materials and Ti to Al interfaces obtained between the CA-SIR and R-FBP frameworks can be attributed to their propagation of scattering effects in image reconstruction. At a large flat interface, like that of the phantom between the Ti to Al half cylinders, the superposition of scattering along its length when the beam direction is either parallel, or near parallel, results in increased levels of localised noise in the recorded images. When the ramp filter is applied during the FBP process to sinogram slices obtained in such an orientation to the Ti to Al interface these high noise region are interpreted as multiple interfaces \cite{Bushburg} and results in an ``averaged'' blurred interface like that seen in the R-FBP recovered estimate of Figure \ref{fig:1}. However in the case of the CA-SIR framework it can be seen that even with one eighth the number of projections it is able to achieve a LSF between the Ti to Al interface on par with R-FBP and decreases, indicating a better interface definition, with increased number of projections. This is due to the fact that SIR frameworks were developed to specifically account from the impact of noise and, when an appropriate system noise model is integrated into the forward projector, these effects can be suppressed proportional to the number of object projections utilised in image reconstruction.

\begin{figure}[tb]
  \centering
  \includegraphics[width=1\textwidth, trim = {45 0 45 0},clip]{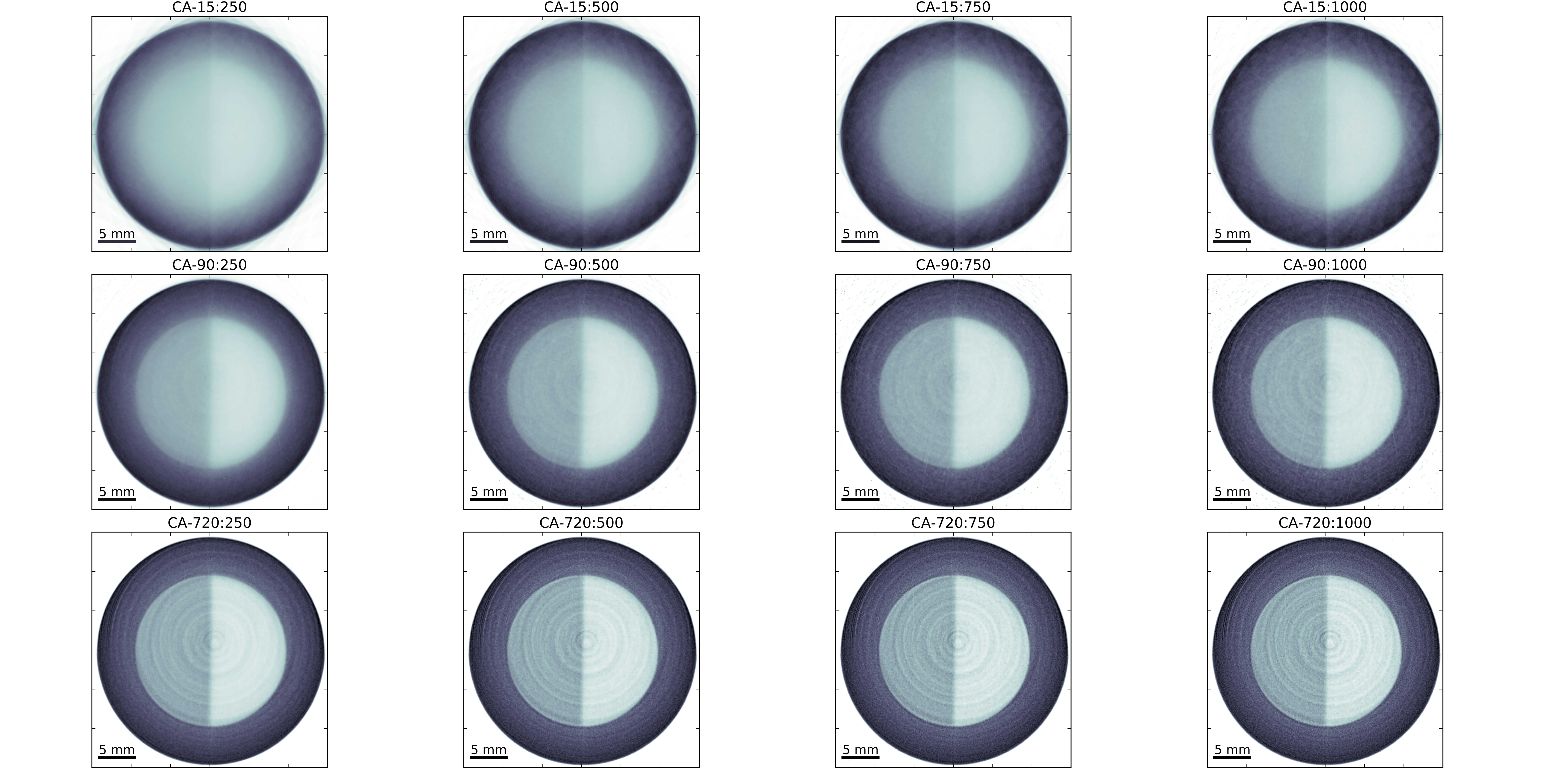}
  \caption{Recovered object neutron linear attenuation coefficient estimate slices from the centre of the imaged phantom for CA-SIR 15, 90 and 720 object projections at iteration numbers 250, 500, 750 and 1000. Here each recovered slice's colour scale spans from zero to the maximum recovered linear attenuation coefficient.}
  \label{fig:6}
\end{figure}

Within this work a total of 1000 iterations was implement for all projection number trials with the CA-SIR framework. This number was selected as an example value, however in reality the number of iterations required for satisfactory data convergence can differ significantly depending on the imaging system and object of study \cite{Lange1,Fessler2000}. To illustrate that in this work sufficient data convergence occurred Figure \ref{fig:6} presents the 250, 500, 750 and 1000 iteration central object slice recovered image estimates with the CA-SIR framework for 15, 90 and 720 projections. Visual assessment of these recovered object estimates illustrate two main points: 1) all three different projection number trials reached convergence by 750 projections and 2) that the rate of convergence with the CA-SIR is proportional to the number of utilised projections. 

At time of this publication the developed CA-SIR framework has been distributed for use at two different neutron imaging facilities across the globe: DINGO, ANSTO, Australia \cite{DINGO1}, and FISH, TUDelft, The Netherlands \cite{Zhou2018}. It will be further tested to explore the possible increase in object throughput at these imaging stations for a variety of different samples. In addition further refinement to this work is planned to upgrade it to a full-quantitative IR framework through the addition of physical processes such as beam hardening, scattering and spectrum dependent detector response \cite{Nuyts2013}. Its is expected that the on-going testing, refinement and redistribution of this CA-SIR framework to neutron imaging facilities such as these will help to increase the accessibility of neutron tomography to the wider scientific community.

\section{Conclusion}
\label{section:C}
A Convex Algorithm Statistical Image Reconstruction (CA-SIR) framework applicable to the geometry of most Neutron CT instruments was developed with the aim of obtaining similar imaging quality to conventional Ramp Filtered Back-Projection via the inverse Radon transform (R-FBP), from a lower number of measured projections to increase object throughput. These two frameworks were applied to a tomographic scan of a cylindrical phantom made up of two half rods of Aluminium (Al) and Titanium (Ti) inserted inside a Stainless Steel (S.S.) tube obtained with the neutron radiography instrument DINGO at the OPAL research reactor (ANSTO, Australia). Comparison of these two IR framework was undertaken both qualitatively and quantitatively via three figures of merit, Signal to Noise Ratio (SNR), Contrast and Line Spread Function (LSF), to discover that the CA-SIR framework is capable of obtaining image estimates of similar quality with respect to R-FBP using only 12.5\% the number of projections. Based on this results CA-SIR has the potential to increase object throughput at neutron imaging facilities such as DINGO, ANSTO, Australia eight-fold.

\section*{Acknowledgements}

We acknowledge the funding of the Australian Nuclear Science and Technology Organisation through the Commissioning Proposal ID 3795. This work was also supported by the Multi-modal Australian ScienceS Imaging and Visualization Environment (MASSIVE) (http://www.massive.org.au).

\end{document}